# Productivity tools to study constrained motion: electrician's approach to mechanician's problem

V.V.Tregub

Department of Physics, Novosibirsk State University
Pirogova 2, Novosibirsk, Russia
vasilich@tregub.ru

**Abstract**. An application design is offered, which students of physics can use when authoring a solver for mechanical systems with constraints.

Authoring a solver application to compute motion of mechanical systems with constraints could be a useful learning experience for students of physics. In such a project, mechanics is closely interwoven with numerical methods and linear algebra. Successful completion of the project will help students to develop a holistic view on university curriculum content. However, teaching assistants may be reserved about assigning this project to non-CS students, for it is all but an exercise in coding of a pure numerical analysis problem. To be handy as a research tool, the solver must have well-designed user interface. In particular, the solver application is expected to provide adequate input/output facilities – even if the only users of the application are the students and their instructors.

As for graphic output, novel class libraries make plotting from code easier these days; another option is use MATLAB, Octave, or Python. In comparison, implementing graphic input might be a daunted task. For educational purposes, I offer to "re-use" the netlist concept from a realm of electronic device simulators. The configuration of a constrained mechanical system with given initial conditions can be captured in the form of a "chainlist", short for a "kinematic chain list", similar in design and purpose to electrical circuit netlists, and afterwards entered into the solver application.

Consider a project in which a student is going to study constrained motion. The student decides on authoring an application that solves the equations of motion for a mechanical system with constraints

$$\begin{aligned} M\ddot{q} + \Phi_q^T\lambda &= F \\ \Phi(q,\dot{q},t) &= 0 \end{aligned} \qquad (1)$$

where $M$ denotes the mass matrix, $q$ is generalized coordinates, $\Phi$ are constraints, $\Phi_q$ is the Jacobian of constraints, $\lambda$ is Lagrange multipliers, $F$ is a vector of external (not imposed by constraints) generalized forces.

Let the student start with only holonomic constraints and gravity as the only external force. In this case, the constraint equations depend explicitly only on generalized coordinates. The system (1) is classified as a DAE (differential-algebraic equations) system of (differentiation) index 3. The brute force method to numerically solve this system results in solving *n* nonlinear equations each time step; *n* is the constraint count.

Alternatively, one may differentiate the constraints and reduce the index. Taking a second derivative of the constraints, we obtain an index 1 DAE system:

$$M\ddot{q} \; + \; \Phi_q^T\lambda = F$$

$$\Phi_q\ddot{q} \; = -\left(\frac{\partial(\Phi_q\dot{q})}{\partial q}\right) - 2\dot{\Phi}_q\dot{q} - \ddot{\Phi} \tag{2}$$

To numerically solve ODEs of this system, we enter generalized velocities $v$ into the equations and solve the system analytically to explicitly express the derivatives of variables $q, v$. First re-write system (2) in matrix form:

$$\begin{bmatrix} M & \Phi_q^T \\ \Phi_q & 0 \end{bmatrix}\begin{bmatrix} \ddot{q} \\ \lambda \end{bmatrix} = \begin{bmatrix} F \\ b \end{bmatrix} \tag{3}$$

where $b = -(\Phi_q\dot{q})_q\dot{q} - 2\dot{\Phi}_q\dot{q} - \ddot{\Phi}$ ; or, if $\dot{\Phi} = 0$, $b = -(\Phi_q\dot{q})_q\dot{q}$.

Udwadia and Kalaba take the first derivative of nonholonomic constraints and the second derivative of holonomic constraints w.r.t. time and write down the resulting equations in the form

$$A\ddot{q} \; = \; b$$

We will denote the Jacobian $\Phi_q$ of holonomic constraints $\Phi(q) = 0$ with the symbol $A$.

When solving the system (3) with respect to $[\ddot{q} \quad \lambda]$, one may notice a zero sub-matrix in the bottom right corner of the coefficient matrix of system (3) and use a “natural” partitioning suggested from this observation:

$$\ddot{q} = \; M^{-1}F + M^{-1}A^T(AM^{-1}A^T)^{-1}(b - AM^{-1}F)$$

$$\lambda = M^{-1}A^T(AM^{-1}A^T)^{-1}(b - AM^{-1}F)$$

For this solution to be valid, matrices $M$ and $AM^{-1}A^T$ must be invertible. This requirement imposes certain conditions on the mass matrix and on the type and quantity of constraints. Under these conditions, $B \; = \; AM^{-½}$ is full row rank and we write:

$$\ddot{q} = \; M^{-1}F + M^{-½}B^T(BB^T)^{-1}(b - AM^{-1}F)$$

F.E.Udwadia and R.Kalaba [1] use Gauss’ principle of least constraint to find generalized accelerations $\ddot{q}$

$$\ddot{q} = \; M^{-1}F + M^{-½}B^+(b - AM^{-1}F)$$

The superscript $+$ denotes Moore-Penrose pseudoinverse operation on matrix. The formula is also valid for rank deficient matrices $B$. For full row rank matrix $B$, the Moore-Penrose pseudoinverse can be computed with the explicit formula $B^+ = B^T(BB^T)^{-1}$, and the expression obtained through matrix partitioning coincides with the Udwadia-Kalaba equations.

Then, add the variable $v \; = \; \dot{q}$. The index 1 DAE system is now written in the form

$$\dot{q} \; = v$$

$$\dot{v} = M^{-1}F + M^{-½}(AM^{-½})^+(b - AM^{-1}F) \tag{4}$$

which is known as the Udwadia-Kalaba equations of constrained motion. The Udwadia-Kalaba equations cover the case of rank deficient matrix $B = AM^{-1/2}$ and can be generalized further to cover the case of singular mass matrix $M$ [2].

The initial conditions give the generalized coordinates and the first derivatives thereof at $t = 0$

$$q(0) = x_0; \dot{q}(0) = v_0 \qquad (5)$$

with the vector $v_0$ also satisfying the constraint conditions derived from $d\Phi/dt = 0$.

The student is going to use the system (4) and the initial conditions (5) and numerically solve a certain class of constrained motion problems. The mechanical system graph (mechanism design) defines the number of rows and columns in matrix $A$ and gives the rules to fill in matrix $A$ and vector $b$ with numerical values. The numerical values of matrix/vector elements depend on the mechanical system configuration (positions and velocities of the mechanism parts).

The following examples show how one could use the “chainlist” description to pass mechanical system data into the solver application. A pendulum moving in an x-y plane, the y axis pointing upward, motion starts from a horizontal position of the pendulum, would be defined with the two lines of text (an asterisk marks a comment line):

```
* 1st Pendulum
B1 (1,0) 1
C1 B1 (0,0)
```

Non-comment lines are sequences of space-separated tokens. I will respect the convention of SPICE, which reserves the first line for the netlist title. In the non-comment lines of the 1st Pendulum chainlist, the first token is the entity name. The first letter of the entity name specifies the entity type. The first non-comment line reads: Body #1, starting position = (1,0), mass = 1. The next line reads: Constraint #1 (a mass-less rod) connects Body #1 (a pendulum bob) to a pivot anchored at (0,0). The rod length need not be given explicitly, for it can be calculated with the initial conditions and anchor coordinates. Appendix A gives a rationale behind choosing this format.

This description lacks data on external forces. For now, we consider only gravity as the source of external (w.r.t. the mechanical system) forces. The acceleration of gravity is a “global scope” variable. We can default it, for example, to $g = 9.81$ .

Let $q = [x \quad y]^T$ . The constraint equation for this configuration is $x^2 + y^2 = (x(0))^2 + (y(0))^2 = 1$ . To put a constraint equation in the form $A\ddot{q} = b$, as the equation of Udwadia and Kalaba requires, we take the second time derivative of the original constraint equation and obtain

$$A = [2x \quad 2y] ; \;\; b = -2({v_x}^2 + {v_y}^2)$$

We have one constraint and two coordinates; therefore the matrix $A$ has one row and two columns. The Moore-Penrose pseudoinverse operation on non-zero one-row matrix $(AM^{-1/2})$ transposes it and divides each element of this matrix-vector by the sum of its squared elements. We come to a system of four ODEs in four variables $x, y, v_x, v_y$ . An explicit method like midpoint or RK4 (with an appropriate time step) can be used to produce an approximate solution. The solution suffers known troubles of explicit methods, adjoined with an "accumulated constraint drift" trouble. We will return to it later.

The next four-line (not counting a title line) chainlist describes a double pendulum:

```
* Double Pendulum
B1 (0,-1) 1
B2 (1,-1) 1
C1 B1 (0,0)
C2 B2 B1
```

The last line defines a constraint imposed by a mass-less rod that connects bodies B2 and B1. This time, $q = [x_1 \quad y_1 \quad x_2 \quad y_2]^T$. We have two constraint equations for this double pendulum

$$x_1{}^2 + y_1{}^2 \;=\; 1$$

$$(x_2 - x_1)^2 + (y_2 - y_1)^2 \;=\; 1$$

which give the constraint matrix and vector to be used with the equation of constrained motion:

$$A = \begin{bmatrix} 2x_1 & 2y_1 & 0 & 0 \\ 2(x_1 - x_2) & 2(y_1 - y_2) & 2(x_2 - x_1) & 2(y_2 - y_1) \end{bmatrix}$$

$$b = \begin{bmatrix} -2(v_{x1}{}^2 + v_{y1}{}^2) & -2((v_{x1} - v_{x2})^2 + (v_{y1} - v_{y2})^2) \end{bmatrix}$$

One can easily infer the rules to form constraint matrices and implement these rules in a piece of code that will be used to generate the constraint matrices for mechanical systems with any given number of point masses and link-type constraints.

The matrix $(AM^{-½})$ for double pendulum has two rows and we need a procedure or function to calculate a Moore-Penrose pseudoinverse of this matrix. The Moore-Penrose pseudoinverse of $B$ can be calculated using the singular value decomposition of $B = U\Sigma V^{\mathrm{T}}$:

$$B^+ = V\Sigma^+ U^{\mathrm{T}}$$

The pseudoinverse operation on a diagonal matrix $\Sigma$ replaces each element of matrix $\Sigma$ with its reciprocal value, if the element's absolute value is greater than a given threshold value, and leaves it with zero otherwise. The singular value decomposition can be computed with the LAPACK function DGESVD. To hide complexities of calling FORTRAN-style LAPACK routines, the instructor may provide students with a static library containing the procedure $pinv([M], tolerance)$.

The Moore Penrose pseudoinverse deserves a special treatment in physics/engineering curricula. The pseudoinverse term of (4) defines force of constraint, and therefore has a clear physical meaning. Later we will see how students may grasp a deeper understanding of constrained motion with an experience gained at implementing this operation in code.

To implement the Moore Penrose pseudoinverse operation in code, a beginning programmer can start with the Greville algorithm [3], p.234. This algorithm gives a compact code of a finite recursive procedure with the number of iteration steps equal to the number of source matrix rows (see Appendix B). It is an apt exercise in coding of linear algebra problems to write the Greville algorithm implementation, first for the two-row constraint matrix of double pendulum mechanism and then for multiple-row matrices.

The next step in project development is writing code to process chainlists of arbitrary mechanisms built with lumped masses and link constraints. Each link-type constraint fixes a distance either between given two bodies or between a body and an anchored pivot. First we parse the chainlist and build two collections, a body collection and a constraint collection. The elements of the collections are objects. The elements of the Body collection are instances of class Body. Class Body has fields for body name, mass, coordinates, and velocities. The elements of the Constraint collection are instances of class Link. Class Link has fields for constraint name, for participating bodies Body1 and Body2, for anchor coordinates, and for link length. If the constraint links a body and an anchored pivot, Body2 field is set null. Or alternatively one can have separate data structures for two-body and fixed-pivot link constraints. Later we will add data structures for other type constraints. Both collections are filled in while parsing the input chainlist. The parsing procedure must be called before the simulation loop starts. A simulation procedure that advances a mechanical system configuration one step in time calls the procedure that returns right hand sides for equations of constrained motion. An outline of the solver application:

```
Declarations:
    Classes Body, Link, Slider, ...;
    Collections Bodies, Constraints, ...;
    (global) var gravity, tolerance, timestep, plotmode, ...;
    Array DataToOutput, ...;
    ...

Procedure Main /*entry point of the solver application*/:
    Parse_Chainlist;
    for t=0; t<stop; t+=timestep
        Simulate;
        if PLOTMODE==ANIMATE
            Redraw mechanical system configuration
            synchronously with animation timer ticks;
    plot for start<=t<stop output arrays saved during simulation;

Procedure Parse_Chainlist:
    Read Chainlist data from file or from GUI multiline textbox;
    for_each chainlistLine in Chainlist.Lines
        Split chainlistLine into tokens;
        Analyze tokens and push object
        into Body or Constraint collection
        OR
        Parse chainlist "command";

Procedure Simulate:
    switch INTEGRATOR
        case EULER
            RHS = Compute_RHS;
            for_each Body B in Bodies
```

```
            B.coords += RHS.rhsq * deltat;
            B.velocity += RHS.rhsv * deltat;
        case MIDPOINT (HEUN, RK4, ...)
            ...
    push output data to DataToOutput lists;

Procedure Compute_RHS:
    /* procedure computes finite difference approx
    * for both dq/dt and dv/dt*/
    for_each Constraint C in Constraints
        switch C.Type
            case Link
                use constraint rules for link constraint
                to fill in A, b;
    /* F is external force */
    /* mpinv is a Moore-Penrose pseudoinverse procedure */
    rhsv = M^(-1)*F + M^(-½) * mpinv( A*M^(-½) ) * (b – A*M^(-1)*F);
    rhsq = v; /* will change with project advancement */
    return { rhsq , rhsv };
```

Some details are given in the Appendix C. The completed code can be tested with the chainlists of Appendix D. Even with this "version 0.1" code, the solver may be used to demonstrate a number of numerical artifacts, which plague more advanced numerical methods as well. For example, "shortcuts" in configuration space off the manifold happen when in one time step of numerical simulation the mechanism leaps over to a configuration, which, for continuous motion, is only reachable via much longer route that lies entirely in the manifold. The student is expected to identify and explain these artifacts.

For a deeper grasp of the chainlist concept, have a look at a 12-link chain chainlist (sorry harmless terminology collision):

```
* 12 Link Chain
B1 (0.2,0) 1
B2 (0.4,0) 1
* add nine similar lines for B3,...,B11
B12 (2.4,0) 1
C1 B1 (0,0)
C2 B1 B2
* add nine similar lines for C3,...,C11
C12 B11 B12
.tran 0 10 1E-3
.options mpitol=1.0E-12 gravity=1
```

If the first token in a line starts with a period, it is a chainlist "command". The command `.tran` with its arguments directs the solver application to do a 10 sec simulation with a timestep of 0.001. The command `.options` sets values of solver parameters like `mpitol` and of global variables like `gravity`. Parameter `mpitol` is a tolerance of the Moore Penrose pseudoinverse algorithm. Alternatively, the programmer of GUI solver application would have to provide user controls to enter simulation parameters.

Moreover, it is convenient to store simulation parameters in a common data source with the mechanism simulated. The chainlist concept wins a point.

The set of chainlist entities is readily extensible to include other type constraints, for example, a prismatic joint for the generic slider-crank mechanism put in motion by gravity:

```
* *Slider Crank Mechanism*
* Crank mass
B1 (-0.1,1) 1
* Slider mass
B2 (-0.2,0) 1
* Crank constraint
C1 B1 (0,0)
* Connecting rod
C2 B1 B2
* Slider constraint
S1 B2 LINE(0,1,0)
.tran 0 16 1E-3
.plot CONSTRAINTS(C1,C2,S1)
.options gravity=1 FRAME=0.4s
*+ PLOTMODE=ANIMATEWITHSTROBO
```

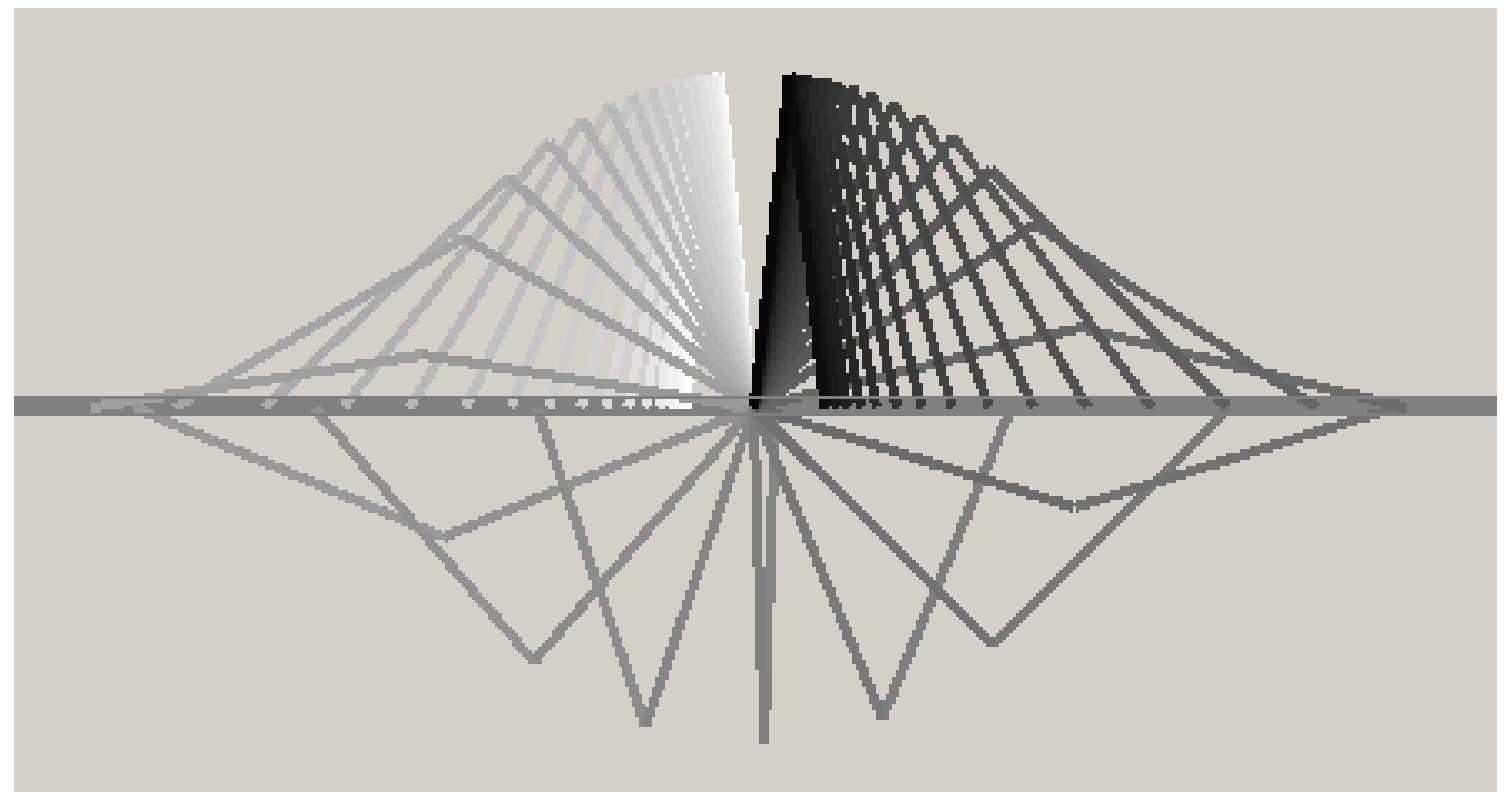

Crank mass B1 is a lumped mass of the travelling spherical joint. Slider constraint S1 constrains body B2 motion to a line `LINE(k1,k2,konst)`. The line equation is $k1 \cdot x + k2 \cdot y = konst$. With $k1 = 0$, $k2 = 1$, $konst = 0$, the line is an x axis. To make parsing easier, the student can implement a syntax like `S[lider] B[ody] k1 k2 konst`. The code that fills in constraint matrix A and vector b must include now the rules both for the link constraint and for the slider constraint.

After the style of SPICE netlists, which may include subcircuit constructs, chainlists may include "subsystems", making the description of mechanism with repeating subsystems more concise and readable:

```
* Lazy tongs
.submachine Tong BA, BB, BC, BD, (x0,y0)
BA (x0-1,y0-1) 1
BC (x0+1,y0+1) 1
BB (x0-1,y0+1) 1
BD (x0+1,y0-1) 1
C1 BA (x0,y0)
C2 BB (x0,y0)
C3 BC (x0,y0)
C4 BD (x0,y0)
C5 BA BC
C6 BB BD
.end Tong
X1 Tong B1, B2, B3, B4, (0,0)
X2 Tong B3, B4, B5, B6
X3 Tong B5, B6, B7, B8
```

Parsing this list may be outside a beginning programmer's reach, as well as implementing other useful language constructs that might be invented for chainlists. I just mention it as a perspective direction of the chainlist "language" development. The extensibility feature wins another point for the chainlist concept.

If one uses Greville's algorithm to code a pseudoinverse procedure as given in Appendix B, the simulation running with the above 12-link chain chainlist completely fails before the simulated chain hits the lowest part of its trajectory in its second swing. Even without running the application in debug mode, one can easily show that it is the pseudoinverse procedure that fails. This failure is not easily amenable to tinkering with mpitol parameter. But change the order from B1,...,B12, in which bodies appear in the 12-link chain chainlist, to B12,...,B1, and the simulation runs smoothly. Since the failure is easily cured with the matrix column reordering, we have clear indication on numerical instability of the pseudoinverse procedure implementation. The chainlist approach wins one more point for being helpful in diagnostics of numerical methods.

Having discovered that the Greville algorithm fails for particular matrices, it is mandate to develop the pseudoinverse procedure implementation using a different algorithm. If matrix $B$ is full row rank, then an equality

$$B^{+} = B^{T}(BB^{T})^{-1} \tag{6}$$

is easily verified to satisfy all the four Penrose equations. In order to invert $BB^{T}$ in a computationally efficient way, first compute Cholesky factorization of $BB^{T} = LL^{T}$, then compute $L^{-1}$ (triangular matrices are easy to invert), and finally, $(BB^{T})^{-1} = (L^{-1})^{T}L^{-1}$.

The pseudoinverse procedure implementing the explicit formula (6) computes matrices of the 12-link chain simulation without numerical instability of the Greville algorithm, but fails on the following problem.

With entities already defined for the chainlist "language", how does one simulate a lever? Actually, we need not define a new type constraint. The lever `((-1,0),(1,0))` with masses 1 and 2 at its ends can be "chainlisted" as follows (the lever fulcrum is at `(0,0)`):

```
* Lever
B1 (-1,0) 1
B2 (1,0) 2
C1 B1 (0,0)
C2 B2 (0,0)
C3 B1 B2
```

Either the lever is unbalanced or balanced (set the B2 mass to 1 to balance this lever), if the pseudoinverse is computed with formula (6), the simulation fails for this chainlist. The simulation with the Greville-based pseudoinverse procedure, on the contrary, succeeds. The reason is the constraint matrix loses rank in this simulation. The student's coder side learns that the pseudoinverse procedure using formula (6) must be augmented to cover the case of rank deficient matrices (see Appendix B). The student's physicist side will hopefully delve into the mechanics of constrained motion and learn that the Udwadia-Kalaba equation holds despite the constraint matrix loses rank.

At this stage of project development, simulation runs are accompanied by an undesirable “accumulated constraint drift”. The link lengths vary with time. For Heun and midpoint integrators the solution behaviors are different, yet both suffer the constraint drift. Doing simulations with chainlists like this (simple pendulum)

```
* pendulum, index1 DAE
B1 (0.001,0.9999995) 1
C1 B1 (0,0)
.tran 0 125 0.1
.plot "Q(B1) PATH(B1)" V(B1) DERIVATIVE(C1)*10
.options gravity=1 DAESYSTEM=INDEX1
```

students may report these behaviors with plots of coordinates-vs-time and velocity-vs-time

```
.options INTEGRATOR=HEUN
```

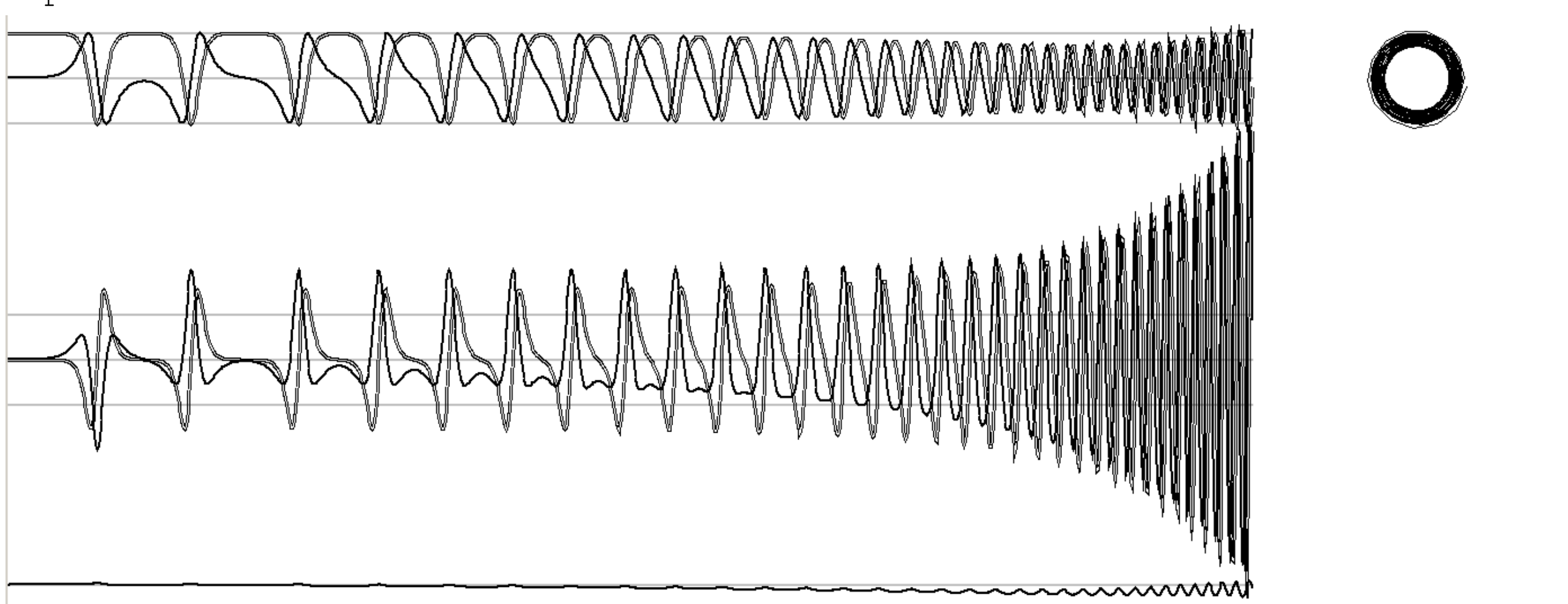

```
.options INTEGRATOR=MIDPOINT
```

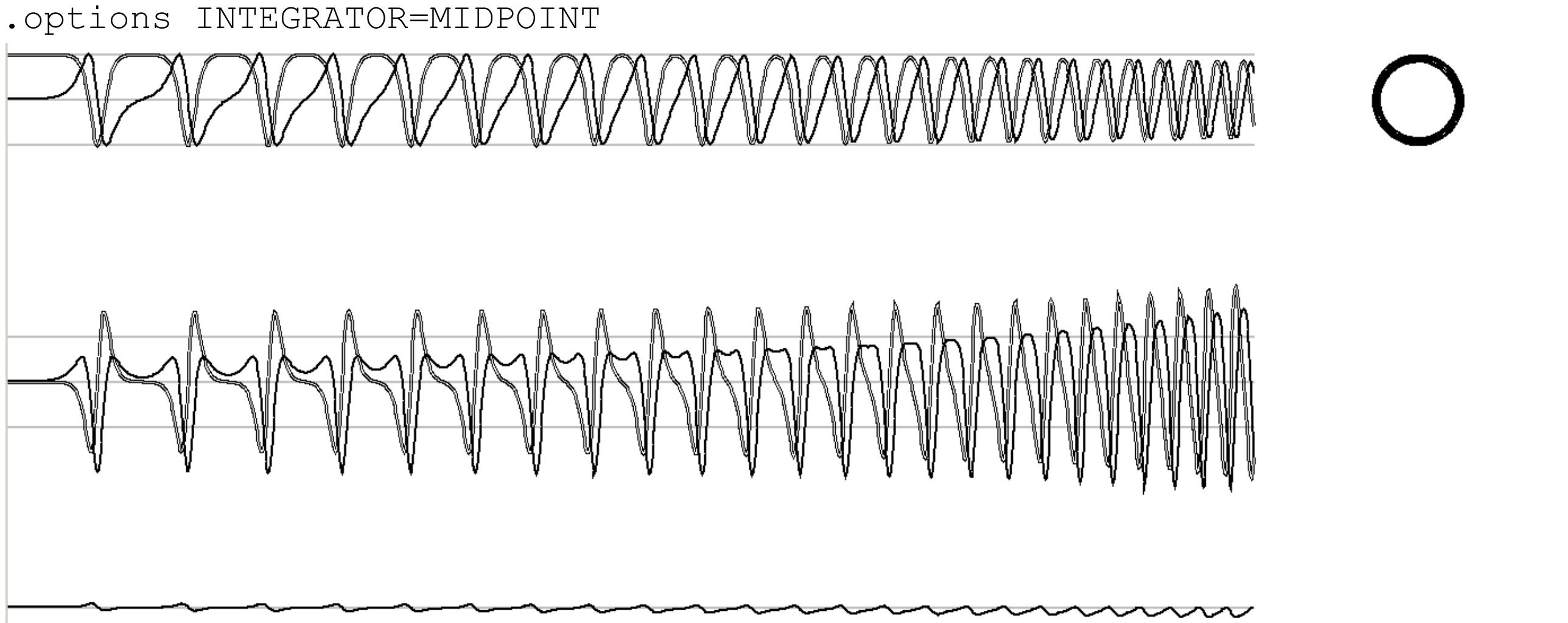

The bottom curve in each figure is an indicator of the velocity constraint violation. The argument `DERIVATIVE(C1)*10` of the `.plot` command instructs the application to plot a full time derivative of constraint C1 times 10. For the simple pendulum, $d\Phi/dt = 2xv_x + 2yv_y$ .

The constraint drift appears because we differentiate the constraint equation, $\Phi(q,t)=0$, with the intention to lower the index of the source index 3 DAE system. Consider a simple pendulum example. Not only the constraint length parameter $l$, but also the equations explicitly binding velocity values, like $2xv_x+2yv_y=0$ for the simple pendulum, are completely omitted from the DAE system after two differentiations. The remaining explicit constraint condition is only imposed on acceleration values. We speculate that if one starts with initial conditions, which fully respect the constraints, and integrates the motion equations EXACTLY, the constraints will propagate as they were. However, in the numerical solution with finite precision and stepsize, numerical errors will accumulate at each time step. The system will move away from the configuration manifold.

Developing the constraint stabilization method, Gear [4] added the velocity constraint into the system (1) explicitly:

$$M\dot{v} + \Phi_q^T\lambda = F$$

$$\dot{q} + \Phi_q^T\mu = v$$

$$\Phi_q(q,t)\dot{q} + \dot{\Phi}(q,t) = 0$$

$$\Phi(q,\dot{q},t) = 0$$

To account for the added constraint equations $d\Phi/dt = 0$, one has to add one more set of Lagrange multipliers, $\mu$. The resulting index 2 DAE system can be solved with Gear's backward differentiation formula and some amount of Newton iterations.

Leveraging Gear's constraint stabilization method, we combine the index 1 and index 2 equations, leave out the position constraint equation, and keep using the explicit methods to solve the resulting system:

$$\begin{gathered} M\dot{v} + \Phi_q^T\lambda = F \\ \Phi_q\dot{v} + (\Phi_q v)_q v + 2\dot{\Phi}_q v + \ddot{\Phi} = 0 \\ \dot{q} + \Phi_q^T\mu = v \\ \Phi_q(q,t)\dot{q} + \dot{\Phi}(q,t) = 0 \end{gathered} \tag{7}$$

Solving for $q,\lambda,v,\mu$ in a way similar to deriving (4), we obtain

$$\begin{gathered} \dot{q} = v + M^{-1/2}(AM^{-1/2})^{+}(b_q - Av) \\ \dot{v} = M^{-1}F + M^{-1/2}(AM^{-1/2})^{+}(b_v - AM^{-1}F) \end{gathered} \tag{8}$$

where $b_v = -(\Phi_q v)_q v - 2\dot{\Phi}_q v - \ddot{\Phi}$ and $b_q = -\dot{\Phi}$.

At this point, the student is expected to report the accuracy of simulations that solve the systems (4) and (8) for a number of mechanisms, and do it in many ways: examine how the solver keeps the constraints, especially when compared with the numerical solution based on a "pure" Udwadia-Kalaba system; compare the numerical simulation and the analytic solution for simple pendulum, etc. The chainlists passed to the solver application may look like this (option DAESYSTEM=INDEX1_2 instructs the solver to use the Gear-Udwadia-Kalaba inspired system (8) )

```
* pendulum, index 1 & 2 mixed eqs
B1 (0.001,0.9999995) 1
C1 B1 (0,0)
.tran 0 125 0.1
.plot "Q(B1) PATH(B1)" V(B1) DERIVATIVE(C1)*10
.options gravity=1 DAESYSTEM=INDEX1_2
```

Having processed this chainlist, the solver outputs the plots:

```
.options INTEGRATOR=HEUN
```

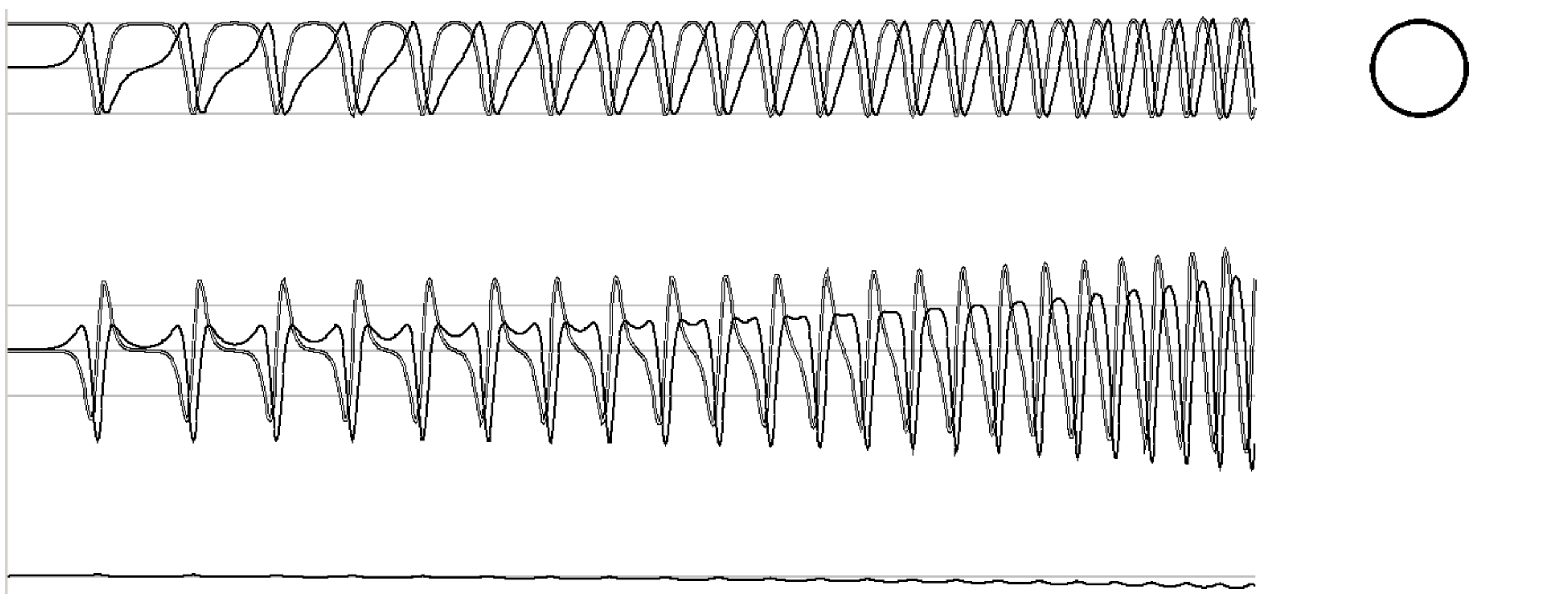

```
.options INTEGRATOR=MIDPOINT
```

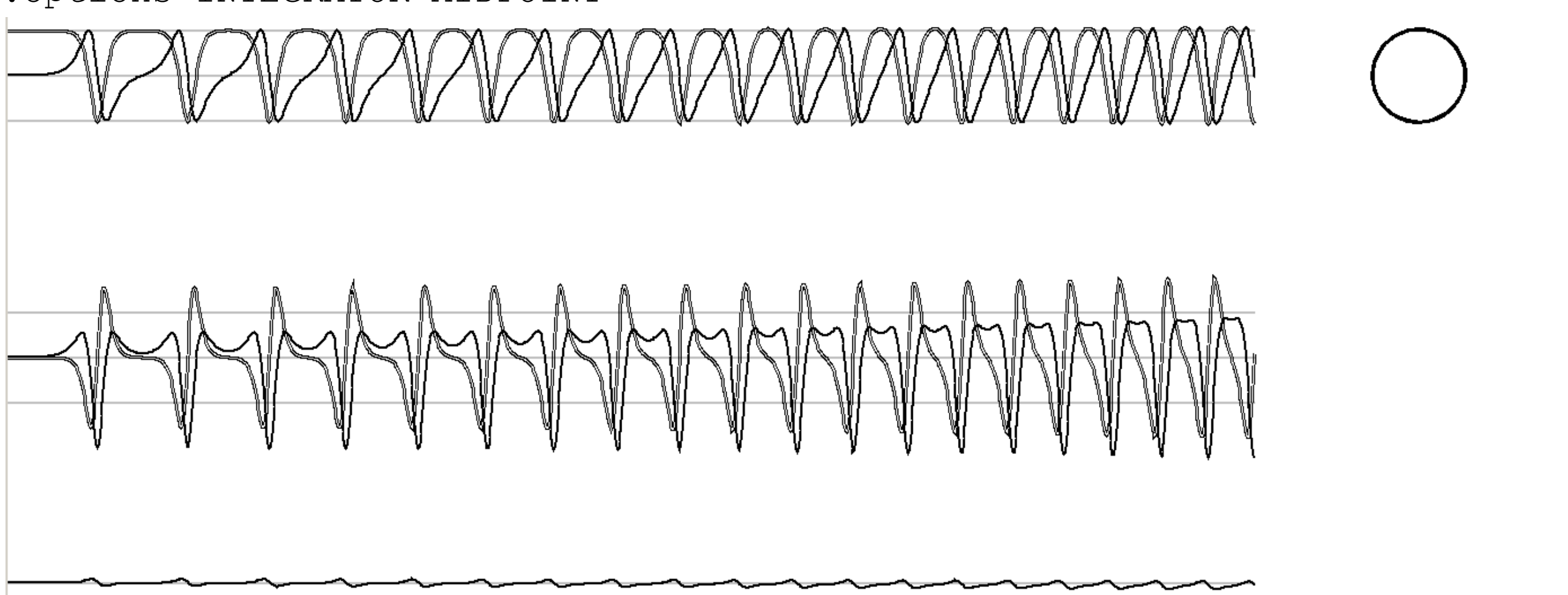

To achieve still more efficient enforcement of constraints, we re-visit system (7) written in the form with separated right hand sides:

$$M\dot{v} + \Phi_q^T \lambda = F$$

$$\Phi_q \dot{v} = b_v$$

$$\dot{q} \; + \; \Phi_q^T \mu = v$$

$$\Phi_q \dot{q} = b_q$$

The equations $\Phi_q \dot{v} = b_v$ and $\Phi_q(q,t)\dot{q} = b_q$, unlike $\Phi(q,t) = 0$ , admit the motion off the manifold. Ideally, this motion is impossible, and appears only as a result of numerical (round-off, discretization etc) errors or inconsistent initial conditions. In numerical solution, we cannot reduce these errors to zero for all times and can only suppress this motion and stay as close to the manifold as possible.

If we require independence from the integration interval, any exact solution of the following system (α, β ≥ 0)

$$\Phi_q \dot{v} = b_v + \alpha \int^t (b_v - \Phi_q \dot{v}) dt$$
$$\Phi_q \dot{q} = b_q + \beta \int^t (b_q - \Phi_q \dot{q}) dt \qquad (9)$$

is also the solution to the system { $\Phi_q \dot{v} = b_v$ ; $\Phi_q(q,t)\dot{q} = b_q$ ; }, and vice versa. For a particular value of the lower limit of integration, the solutions of system (9) approach the solutions of the system { $\Phi_q \dot{v} = b_v$ ; $\Phi_q(q,t)\dot{q} = b_q$ ; } asymptotically. Replacing the constraint equations of system (7) with the equations (9), we introduce an exponential damping of the traverse (w.r.t. the manifold) motion with damping coefficients of $\alpha$ for velocity and $\beta$ for coordinates.

Recalling the definition of $b_q, b_v, \Phi_q$ , we can re-write the equations (9) in the form

$$\Phi_q \dot{v} = b_v - \alpha \dot{\Phi}$$

$$\Phi_q \dot{q} = b_q - \beta \Phi$$

and come up with the explicit equations of index 1_2 DAE system with the forcibly damped off-manifold motion:

$$\dot{q} \; = v + M^{-1/2}(AM^{-1/2})^{+}(b_q - Av - \beta\Phi)$$
$$\dot{v} = M^{-1}F + M^{-1/2}(AM^{-1/2})^{+}(b_v - AM^{-1}F - \alpha\dot{\Phi}) \qquad (10)$$

Using a finite-difference analog of these equations in the solver, with a suitable choice of damping coefficients we enjoy practically exact enforcement of position constraints. Having processed the chainlist with the "natural" choice of values $\alpha = 1/\Delta t, \; \beta = 1/\Delta t$ , which provides critical damping (`DAMPALPHA`, `DAMPBETA` parameters are in units of timestep `deltat`),

```
* pendulum, off-the-manifold motion damped
B1 (0.001,0.9999995) 1
C1 B1 (0,0)
.tran 0 125 0.1
.plot "Q(B1) PATH(B1)" V(B1) DERIVATIVE(C1)*10
.options gravity=1 DAESYSTEM=INDEX1_2_DAMPED DAMPALPHA=1 DAMPBETA=1
```

the solver outputs the plots showing excellent suppression of position constraint drift at any time scale before energy blowup effectively destroys simulation:

```
.options INTEGRATOR=HEUN
```

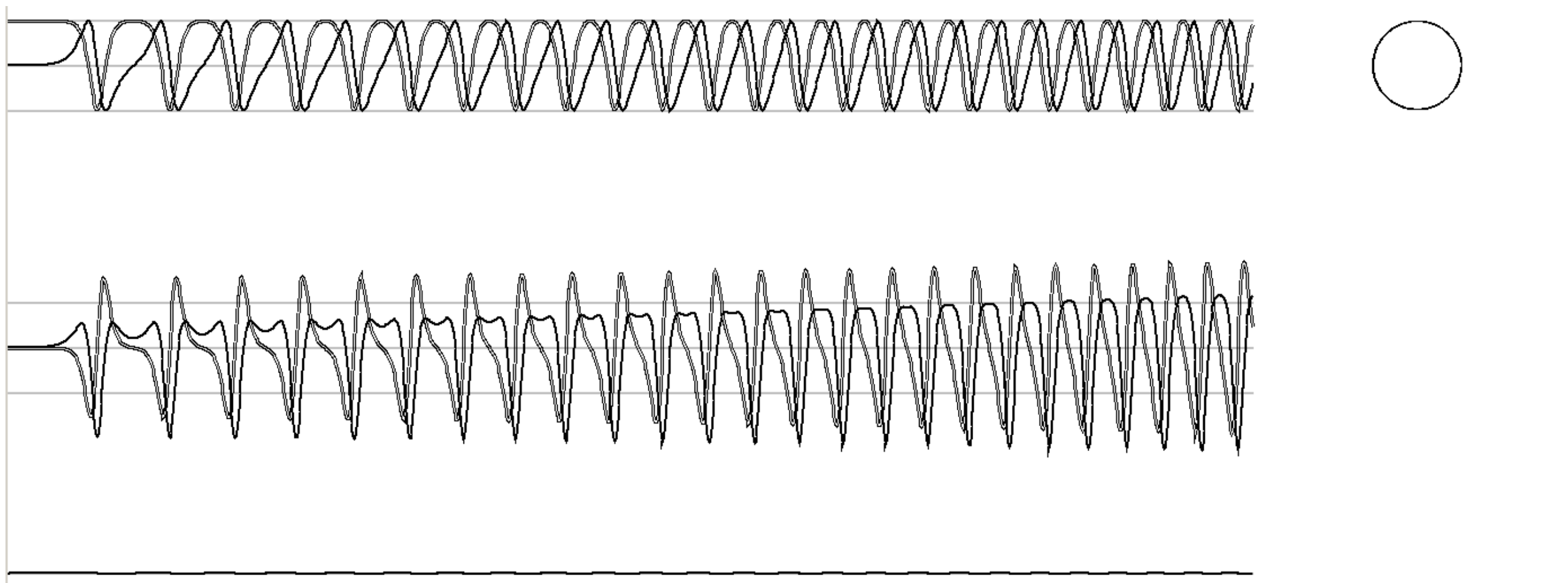

```
.options INTEGRATOR=MIDPOINT
```

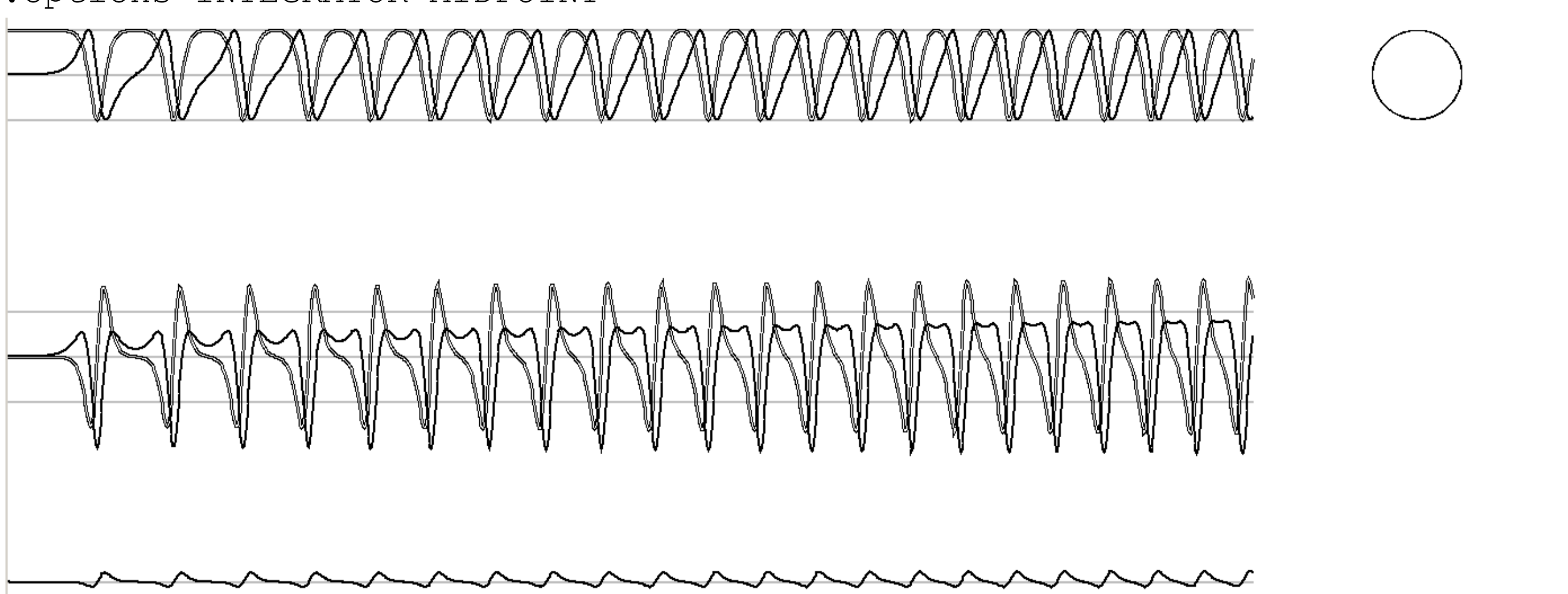

It will be instructive for students to examine the “transients” and the impact of damping factor choice on simulation that starts with inconsistent initial conditions:

```
* pendulum, damped traverse motion, inconsistent ICs
B1 (1,0) 1 IC_VEL=(0,-1) IC_POS=(0.5,0.5)
C1 B1 (0,0)
.tran 0 10 0.05
.plot Q(B1) V(B1) DERIVATIVE(C1) PATH(B1)
.options gravity=1 DAESYSTEM=INDEX1_2_DAMPED DAMPALPHA=1 DAMPBETA=1
```

With the named parameter IC_POS (Initial Conditions for POSition) mentioned on the parameter list of the body instance B1, the first parameter (1,0) is only used to calculate the length of constraint C1. The body B1 starts its motion from a point IC_POS with the velocity IC_VEL (Initial Conditions for VELocity). In my implementation of chainlist parser, unnamed parameters are required, named parameters are optional.

Tackling an energy blowup problem of the above simulations is a further item on the solver development agenda. In general, the Hamiltonian of the constrained mechanical system cannot be separated into kinetic and potential parts. For such systems the explicit symplectic integrators, if ever possible, are difficult to construct. Symplectic integrators and other well-behaved methods for constrained motion problems are presently a hot topic of scientific publications, and implementation of these methods in code requires a serious effort.

As a stage finish, students may implement a toy simulator for constrained mechanical systems without dissipation. With a sort of a predictor-corrector method, the simulator will produce perpetual animations of constrained motion with no noticeable constraint drift. This "predictor-corrector" method conserves energy by design and proceeds as follows. In the beginning of simulation, the total energy E for the initial configuration is computed. For each timestep, in the predictor step the system (10) is solved and the kinetic energy T and the potential energy U are computed. In the corrector step, the velocities are updated to provide for conservation of energy:

```
if (T>T_tol && E>U)
  for (int i = 0; i<bodies; i++)
    Body[i].v *= SQRT((E-U)/T);
```

Students are expected to rationalize this kind of "predictor-corrector" method and to point out its weaknesses with the help of relevant simulations.

## Conclusions

The kinematic chain lists, "chainlists", are a means of description of constrained mechanical system configuration. They can also contain the metadata for the numerical solver application, like simulation interval, timestep, what numerical methods to use, directions for output.

Once the principles of chainlist design are grasped, students and researches can easily use chainlists in their applications and also communicate the simulation scenarios and results in concise, comprehensible form. There is no need for special effort to "standardize" chainlist format. The chainlist authors only need to take care that newly introduced parameter names and "language" constructs be self-explanatory, easy to parse, and to provide necessary comments for first-time use.

There is inevitably a question of defaults. If the parameter value is crucial for simulation, it is a good practice to specify its value, whether it is well-known and generally agreed upon or not. The use of parallel computations or interval arithmetic may be a challenge for specifying these in simulation description, but an alternative is making colleagues to peruse your code. As for computational physics education, inventing communication means is a good exercise for developing students' creativity in physics and engineering science.

## Appendix A

In our time of freely available lexical analyzers and parser generators, it may seem naïve to adopt SPICE netlist syntax for new tasks, all the more for the tasks seemingly unrelated to EDA tasks. More hip would be to develop an XML schema, call it something like KineML, and enjoy readily available XML parsers.

Still, in many ways the netlist approach appears not only adequate, but even promising. It is adequate, because chainlists can be efficiently parsed with a powerful instrument of regular expressions, which can be easily communicated to students. No need to delve into FLEX/Bison or xmllib business. Also, if one agrees to abide by stricter rules of netlist authoring, a simple string splitting (`C++: strtok(...), C#: .Split()`) will do the job of tokenizing and the switch operator can be used to efficiently analyze nearly all chainlist constructs of this paper. Due to simple and terse syntax, chainlists can be viewed and edited with multiline textboxes of moderate sizes that can be easily accommodated within the application window. The syntax is intuitive, easily extensible, and hopefully helps develop students' creativity in engineering science.

The netlist approach holds promise of accommodating "mixed" motion tasks, a term invented apropos on analogy with mixed signal circuit (analog and digital circuit) analysis tasks. Mechanical constraints are considered ideal for the most part of simulation and only in the vicinity of singularities the rigidity, stretching, elasticity etc of constraint materials come into play. The same for collisions: the time evolution of mechanical system is considered to be a series of collision events with a "free" (governed by constrained motion laws) motion in between. The collision phase computations may require use of continuum mechanics methods, etc.

A "chain list" term maybe reminiscent of queries and searches or even hotels and supermarkets, "parts list" would be a more suitable phrase with established meaning. Still, in referring to a mechanical analog of electrical circuit netlists I decided on emphasizing "connectivity" (kinematic chain) rather than bill-of-materials (parts) meaning.

## Appendix B

For mechanisms with lumped masses, mass matrices are diagonal in Cartesian coordinates. Inversion and square root extraction operations have trivial implementations for diagonal matrices with positive diagonal elements.

Generalized Cholesky decomposition for a symmetric positive semi-definite matrix `A[m,m]` (C# code):

```
double tol; // tolerance to pronounce the matrix singular
int r=0;
double[,] L = new double[m,m]; // Cholesky factor
for (int i=0; i<m; i++)
{
  for (int j=i; j<m; j++)
  {
    L[j,r] = A[j,i];
    for (int k=0; k<r; k++)
      L[j,i] -= L[j,k]*L[i,k];
  }
```

```
        if (L[i,r]>tol)
        {
          L[i,r] = Math.Sqrt(L[i,r]);
          if (i<m-1)
            for (int j = i+1; j<m; j++)
              L[j,r] /= L[i,r];
          r++;
        }
      }
```

The programmer should remember that while the Cholesky factor array is declared as `L[m,m]`, in fact it is a matrix `L[m,r]`. The Cholesky factor matrix dimensions are `m*r`.

The generalized Cholesky factor $L$ of the positive semi-definite matrix $A$ is a square root of this matrix, and the generalized Cholesky decomposition can be used to compute square roots of non-diagonal mass matrices.

The reverse-order product of the generalized Cholesky factors, $(L^T L)$, is a symmetric positive definite matrix, which is invertible. This property can be used to find an analog of the formula $B^+ = B^T(BB^T)^{-1}$ for the case of rank deficient matrices. The book Generalized Inverses: Theory and Applications by Ben-Israel and Greville [3] gives a formula $A^+ = A^*(AA^*)^+$ for complex matrix $A$ (Exercise 18 (d), pp. 42-43). For real matrix $B$

$$B^+ = B^T(BB^T)^+ \tag{B.1}$$

The pseudoinverse of a rank deficient matrix is expressed via the pseudoinverse of a symmetric positive semi-definite matrix. The generalized Cholesky decomposition expresses the symmetric positive semi-definite matrix $(BB^T)$ as a product of two full row rank matrices $(LL^T)$ . The first Penrose equation for $(LL^T)^+$, $(LL^T)(LL^T)^+(LL^T) = (LL^T)$, gives a hint how to construct the explicit formula for pseudoinverse computation. If the sought expression starts with $L$ and ends with $L^T$ ( $(LL^T)^+ = L \cdot \ldots \cdot L^T$ ), these outer factors can be paired with neighboring matrices from the Penrose equation to form full-rank square, symmetric positive definite, invertible matrix $(L^T L)$ to the left and to the right of the ellipsis in the following equation

$$(LL^T)(L \cdot \ldots \cdot L^T)(LL^T) = L(L^T L) \cdot \ldots \cdot (L^T L)L^T$$

By inspection, $L(L^T L)^{-1}(L^T L)^{-1}L^T$ satisfies the first Penrose equation. The equality

$$(LL^T)^+ = L(L^T L)^{-1}(L^T L)^{-1}L^T \tag{B.2}$$

can be directly verified to satisfy all four Penrose equations. Combining (B.1) and (B.2), we have an explicit formula for a pseudoinverse of matrix $B$ [6]:

$$B^+ = B^T L(L^T L)^{-1}(L^T L)^{-1}L^T \tag{B.3}$$

where $L$ is the generalized Cholesky factor of $(BB^T)$.

To efficiently invert the matrix $(L^T L)$ , one can use the (classical) Cholesky decomposition $(L^T L) = MM^T$ . Lower triangular square matrix $M[r,r]$ is full rank and can be inverted with a compact piece of C# code:

```
double[,] invM = new double [r, r];
for (int i=r-1; i>=0; i--)
  for (int j=i; j>=0; j--)
    if (i==j)
      invM[j,j] = 1/M[j,j];
    else
      for (int k=j+1; k<=i; k++)
        invM[j,i] -= M[k,j]*invM[k,i]/M[j,j];
```

Greville's algorithm written in pseudocode (array indices as in FORTRAN; central dots denote inner product operation and full stops in the index field denote bound variables; apostrophe denotes matrix transpose ):

```
given A[1:m,1:n]
find AX[1:n,1:m] = Moore-Penrose pseudoinverse of A
mpitol = n*maxdiag(A*AT)*Double.Epsilon;
aa = A[.,1]'·A[.,1]; // inner product of first column with itself
if (aa > mpitol) AX[1,] = A[,1]' / aa; else AX[1,] = 0;
for iter=2; iter<n; iter++
  new d[1:iter] = AX[1:iter,.]·A[.,iter];
  new c[1:m] = A[,iter] - A[,1:.:iter]·d[.];
  cc = c'[.]·c[.]; // inner product of c[] with itself
  new b[1:m];
  if ( cc > mpitol )
    b[] = c[] / cc;
  else
    b[] = d'[.]·AX[1:.:iter,]/(1+(d'[.]·d[.]));
  AX[1:iter,] = AX[1:iter,] - d[1:iter]*b[];
  AX[iter+1,] = b[];
```

## Appendix C

The users of UNIX/Linux OS's have at their disposal distribution-provided software development tools. For some time, the MS Windows users were supposed to use Microsoft Visual Studio or third party tools for software development. With .Net Framework, the Windows users have the C# compiler as part of OS distribution and can compile without Visual Studio installation. The C# compiler is an executable csc.exe in the folder C:\WINDOWS\Microsoft.NET\Framework\vX.X.XXXX. The C# language is similar in syntax to C++ and has multi-dimensional arrays and other constructs useful for scientific computing. The performance of the C# compiler-produced code approaches the performance of code produced by C++ compilers. While weighing C# pros and cons, consider the benefits of using .NET Framework class libraries and easier memory management with the .NET Framework infrastructure.

The following code excerpts can be used to start off the project. Production code must contain a number of elaborations such as processing of exceptions, generation of parsing error reports, etc.

This sample code parses a chainlist lines passed from a multiline textbox `tbChainlist`:

```
  // elsewhere earlier in code:
  public ArrayList Bodies = new ArrayList(); // body collection
  public ArrayList Constraints = new ArrayList();// constraint collection
  /* chainlistLine is like "B1 (0.0,1.0) 1.0" OR ="C1 B1 B2" */
foreach (string chainlistLine in tbChainlist.Lines)
{
  if (chainlistLine[0] == '*') continue; // comment line
  // remove extra chars from parenthesized tokens like ( 0.0, 1.0)
  string currLine = Regex.Replace( s, @"\(.*?\)",
     new delegate(Match m) {
           return Regex.Replace(m.ToString(), @"[\(\)\s]", String.Empty);
         } );
  // tokenize string
  string [] chips = currLine.Split(default(Char[]),
                       StringSplitOptions.RemoveEmptyEntries);
  // chainlist "command" starts with '.'
  if (currLine[0] == '.')
  {
    switch (chips[0])
    {
      case ".tran":
        starttime = Double.Parse(chips[1]);
        stoptime = Double.Parse(chips[2]);
        timestep = Double.Parse(chips[3]);
        break;
      case ".options": /* .print,.plot etc*/
        ...
        break;
    }
    continue;
  }
  // process chainlist entities (Bodies, Constraints)
  switch (currLine[0])
  {
    case 'B':
      string[] pos = chips[1].Split(new Char[] {','});
      Bodies.Add( new Body( chips[0], Double.Parse(pos[0]),
                            Double.Parse(pos[1]),
                            Double.Parse(chips[2]) ) );
      break;
    case 'C':
      Body B1=null;
      Body B2=null;
      foreach (Body B in Bodies)
        if (B.bodyName==chips[1])
```

```
            B1 = B;
        if (B1==null)
          break;
        if ( chips[2].Contains(",") )
        {
          string[] pos = chips[2].Split(new Char[] {','});
          double ancX = Double.Parse(pos[0]);
          double ancY = Double.Parse(pos[1]);
          double len2 = (B1.x-ancX)*(B1.x-ancX)+(B1.y-ancY)*(B1.y-ancY);
          Constraints.Add( new Link( chips[0], B1, ancX, ancY, len2) );
        }
        else
        {
          foreach (Body B in Bodies)
            if (B.bodyName==chips[2])
              B2 = B;
          if (B2==null)
            break;
          double len2 = (B1.x-B2.x)*(B1.x-B2.x)+(B1.y-B2.y)*(B1.y-B2.y);
          Constraints.Add( new Link( chips[0], B1, B2, len2 ) );
        }
        break;
  }
}
```

The code suitable to run on a localized OS should parse numbers with overloaded methods like `Double.Parse("3.14", CultureInfo.InvariantCulture)`.

With a dozen more lines of code, the beginning programmer can do without regexps when parsing simple chainlist constructs, but I could not resist showing off a “syntactic sugar” of C# `delegate` with anonymous function.

To draw a plot of `motion_arr[samples,1]` vs. `motion_arr[samples,0]` with a compound pen, one should write in the form’s OnPaint event handler:

```
    protected override void OnPaint(PaintEventArgs e) {
        Pen compPen = new Pen(Color.Black);
        compPen.Width = 3;
        compPen.CompoundArray = new float[] {0F, 0.333F, 0.667F, 1F};
          GraphicsPath fPlot = new GraphicsPath();
          Point[] pts = new Point[samples];
          for (int i = 0; i < samples; i++)
            pts[i] = new Point((int)(motion_arr[i,0]*1000/range),
                (int)((vertOffset - motion_arr[i,2]/height)*1000));
          fPlot.AddLines(pts);
          e.Graphics.DrawPath(compPen, fPlot);
```

```
        compPen.Dispose();
    }
```

In the study of constrained motion, animations are a useful tool to augment an experience gained at examining plots and phase portraits. A few of Appendix D chainlists produce rather amusing animations. The application can produce data for animation "on the fly", synchronously with animation timer, or it can write configuration data (positions, velocities, accelerations) at each time step into a tracefile. The tracefile can be used later to draw plots or play animations.

The code to write a data record to the tracefile is like this:

```
simulate();
simtime += deltat;
using (StreamWriter sw = File.AppendText(tracefilename))
{
  sw.Write(simtime);
  foreach (Body B in Bodies)
    sw.Write(" {0}({1},{2})", B.bodyName, B.x, B.y);

  sw.WriteLine();

}
```

You can write binary tracefiles using `BinaryWriter`.

The trace file header may contain the chainlist text and other metadata. In data section, the simulator writes current time and mechanical system configuration for each simulation step.

```
HEADER

* * 9 Link Chain
B1 (0.2,0) 1
...
B9 (1.8,0) 1
C1 B1 (0,0)
...
C9 B8 B9
.tran 0 10 1E-3
.options gravity=1 MPI_ALGO=GREVILLE mpitol=1.0E-11
.options DAESYSTEM=INDEX1_2_DAMPED DAMPALPHA=1 DAMPBETA=1
.options INTEGRATOR=MIDPOINT ENERGYCORRECTOR=ON
.print TEXTTRACEFILE="9linkChain.txt" FORMAT="0.###E0" Time Bodies(POS)

DATA SECTION

0 B1(0.2,0) B2(0.4,0) B3(0.6,0) B4(0.8,0) B5(1,0) B6(1.2,0) B7(1.4,0)
B8(1.6,0) B9(1.8,0)
```

```
0.001 B1(0.2,-5E-7) B2(0.4,-5E-7) B3(0.6,-5E-7) B4(0.8,-5E-7) B5(1,-5E-7)
B6(1.2,-5E-7) B7(1.4,-5E-7) B8(1.6,-5E-7) B9(1.8,-5E-7)

0.002 B1(0.2,-2E-6) B2(0.4,-2E-6) B3(0.6,-2E-6) B4(0.8,-2E-6) B5(1,-2E-6)
B6(1.2,-2E-6) B7(1.4,-2E-6) B8(1.6,-2E-6) B9(1.8,-2E-6)

...

1.732 B1(6.676E103,-1.177E104) B2(9.682E103,-8.698E102) B3(2.918E104,
-1.411E104) B4(5.034E89,1.886E89) B5(4.79E89,-2.146E89) B6(5.065E89,
-1.953E89) B7(-1.632E120,-2.702E119) B8(-1.402E118,2.832E120)
B9(2.056E103,4.652E105)

1.733 B1(-6.588E152,1.162E153) B2(-8.247E152,5.602E152) B3(1.627E138,
1.214E137) B4(4.258E166,2.182E134) B5(4.313E166,9.187E165) B6(3.933E166,
6.511E165) B7(-9.421E148,-1.256E134) B8(-9.829E147, 2.167E149)
B9(3.356E134,1.109E136)

1.734 B1(NaN,NaN) B2(NaN,NaN) B3(NaN,NaN) B4(NaN,NaN) B5(NaN,NaN)
B6(NaN,NaN) B7(NaN,NaN) B8(NaN,NaN) B9(NaN,NaN)
```

## Appendix D

For free-hanging chains with great many links, simulations are numerically stable, but greedy for CPU time.

```
* *16 Link Chain*
B1 (0.2,0) 1
...
B16 3.2 0 1
C1 B1 (0,0)
...
C16 B15 B16
.tran 0 10 0.001
.plot CONSTRAINTS(ALL)
.options gravity=1 FRAME=0.4s
*+ PLOTMODE=ANIMATEWITHSTROBO
```

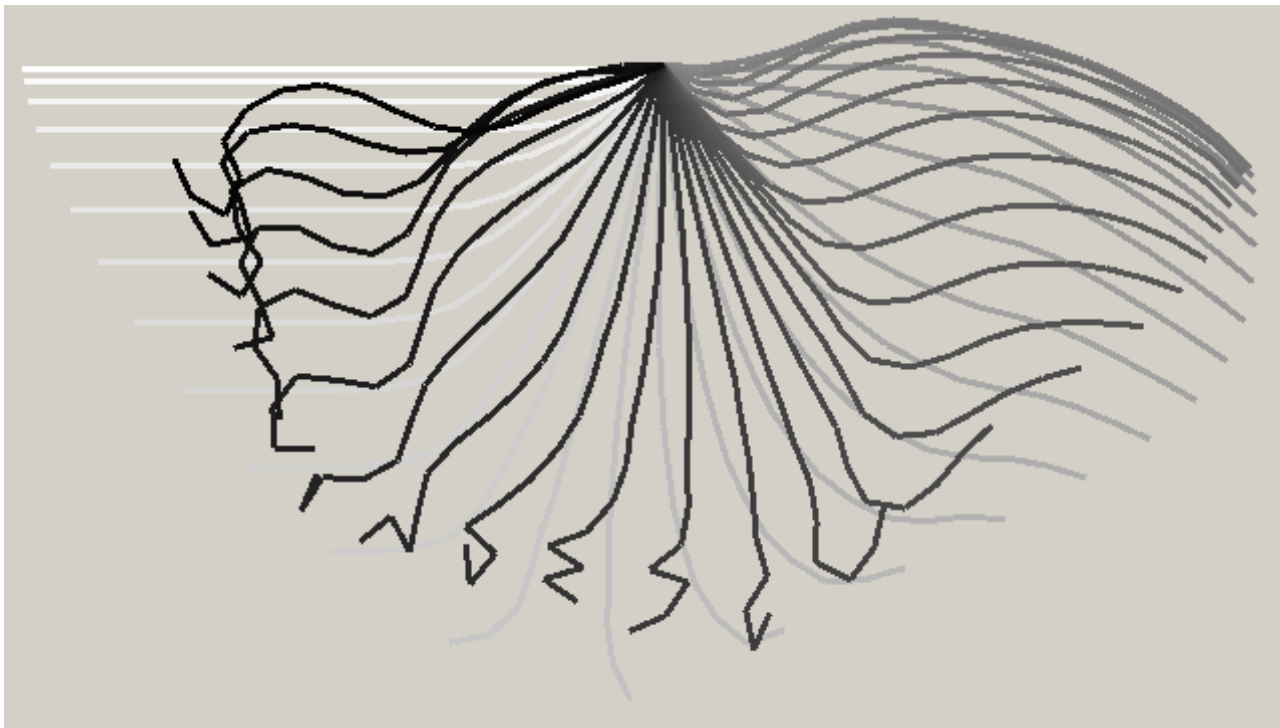

In the following simulation, the last frame is drawn at irregular time interval of 0.057 s following the penultimate, regular frame. The last frame time is 1.857s. At 1.9 s, the configuration is completely destroyed because of singularity.

```
* *Single-Pivot Rhombus Linkage*
B1 (-1,1) 1
B2 (1,1) 1
B3 (0,2) 1
C1 B1 (0,0.0000001)
C2 B2 (0,0.0000001)
C3 B1 B3
C4 B2 B3
.tran 0 2 0.001
.plot CONSTRAINTS(ALL)
.options gravity=1 FRAME=0.1s
*+ PLOTMODE=ANIMATEWITHSTROBO
.options DAESYSTEM=INDEX1_2_DAMPED
*+ DAMPALPHA=1 DAMPBETA=1
.options ENERGYCORRECTOR=OFF
*+ INTEGRATOR=MIDPOINT
```

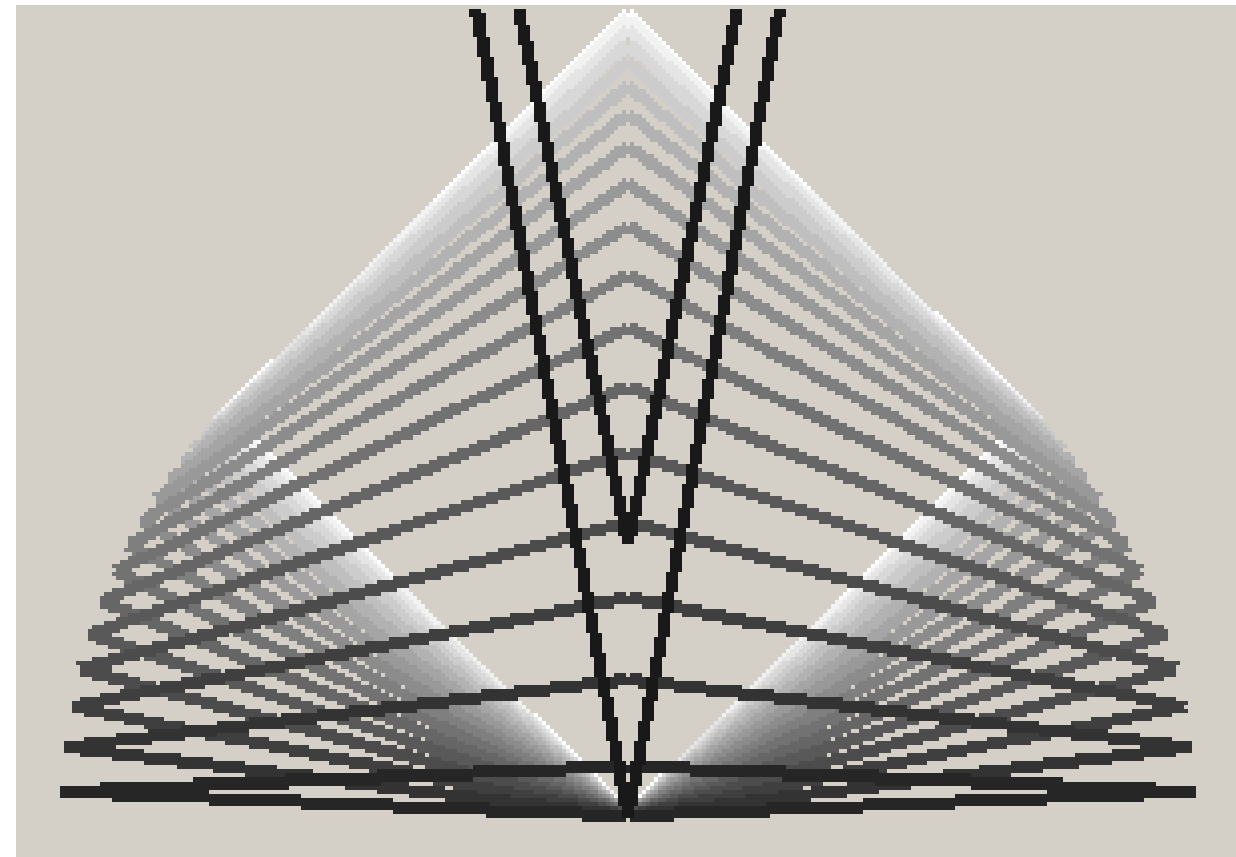

Turning energy corrector on and halving a timestep does repair visual representation, and singularity can only be noticed through tracefile inspection. At 1.8535 s, when the bodies cross zero height, the total energy increases by more than half, and at 1.86 s returns to small fluctuations around its original value.

```
* *Single-Pivot Rhombus Linkage*
B1 (-1,1) 1
B2 (1,1) 1
B3 (0,2) 1
C1 B1 (0,0.0000001)
C2 B2 (0,0.0000001)
C3 B1 B3
C4 B2 B3
.tran 0 2.5 0.0005
.plot CONSTRAINTS(ALL)
.print TEXTTRACEFILE=CONSOLE
*+ Time Etotal T U Bodies(POS)
.options gravity=1 FRAME=0.1s
*+ PLOTMODE=ANIMATEWITHSTROBO
.options DAESYSTEM=INDEX1_2_DAMPED
*+ DAMPALPHA=1 DAMPBETA=1
.options ENERGYCORRECTOR=ON
*+ INTEGRATOR=MIDPOINT
```

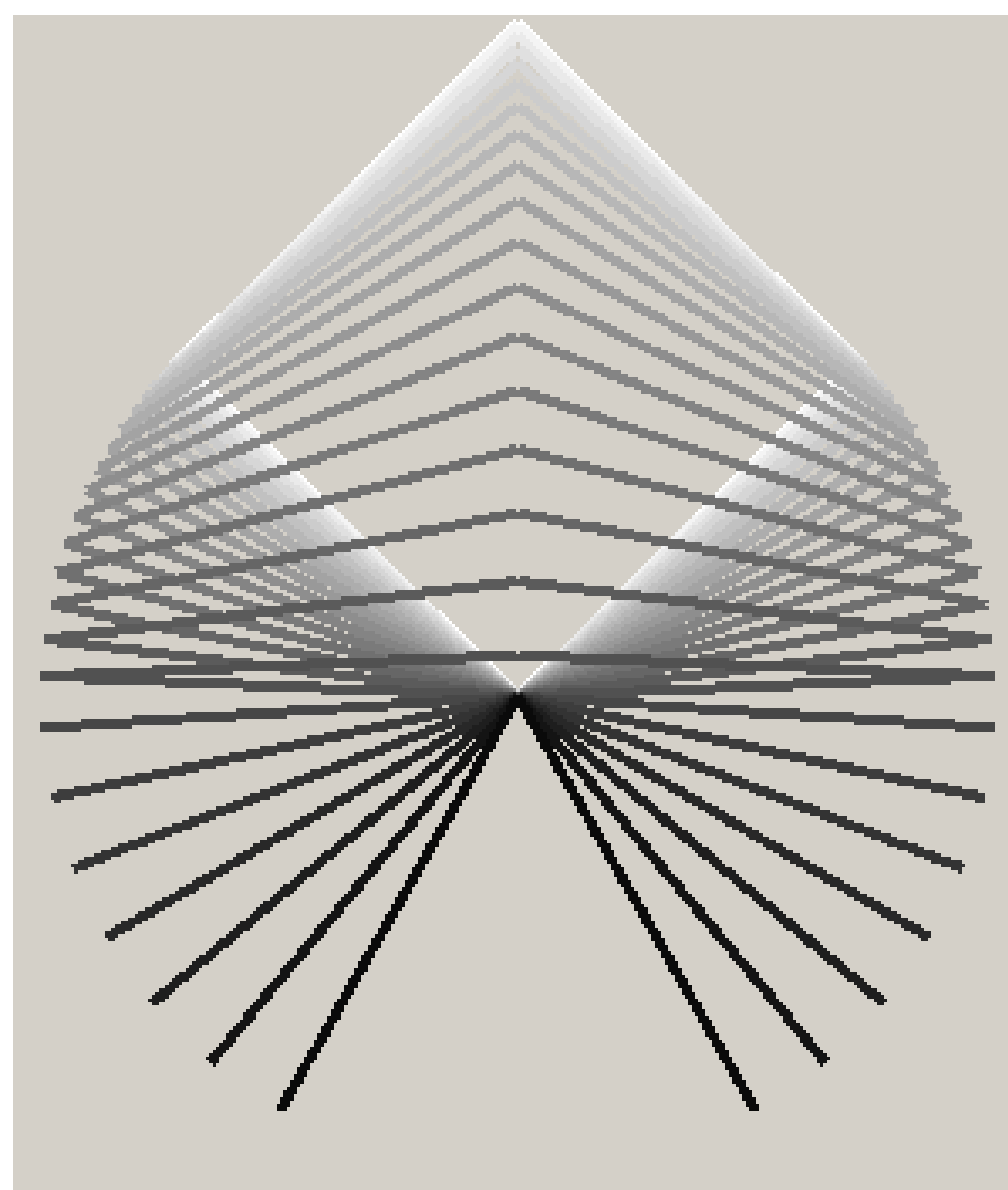

A five-link chain with endpoints anchored at equal heights oscillates with alternate phases of motion until numerical error destroys symmetry:

```
* * 5 links, anchored endpoints *
B1 (-1,1) 1
B2 (1,1) 1
B3 (0,2) 1
C1 B1 (-2,0)
C2 B1 B3
C3 B2 B3
C4 B2 (2,0)
.tran 0 8 0.001
.plot CONSTRAINTS(ALL)
.options gravity=1 FRAME=0.4s
*+ PLOTMODE=ANIMATEWITHSTROBO
```

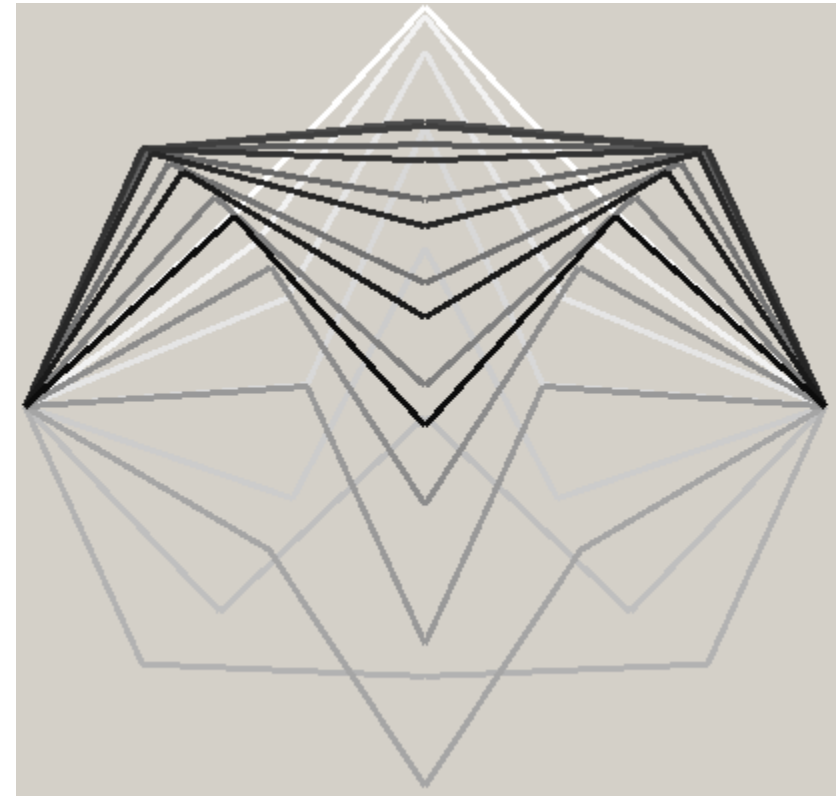

"Inverted" rotation of a link that connects two bodies, constrained also by two slider constraints positioned at right angles. Gravity is zero:

```
* Crosshair Slider Constraints
B1 (0,0) 1 IC_VEL=(1,0)
B2 (0,1) 1
C1 B1 B2
S1 B1 LINE(0,1,0)
S2 B2 LINE(1,0,0)
.tran 0 6.28 0.001
.plot CONSTRAINTS(C1)
.options gravity=0 FRAME=0.2s
*+ PLOTMODE=ANIMATEWITHSTROBO
```

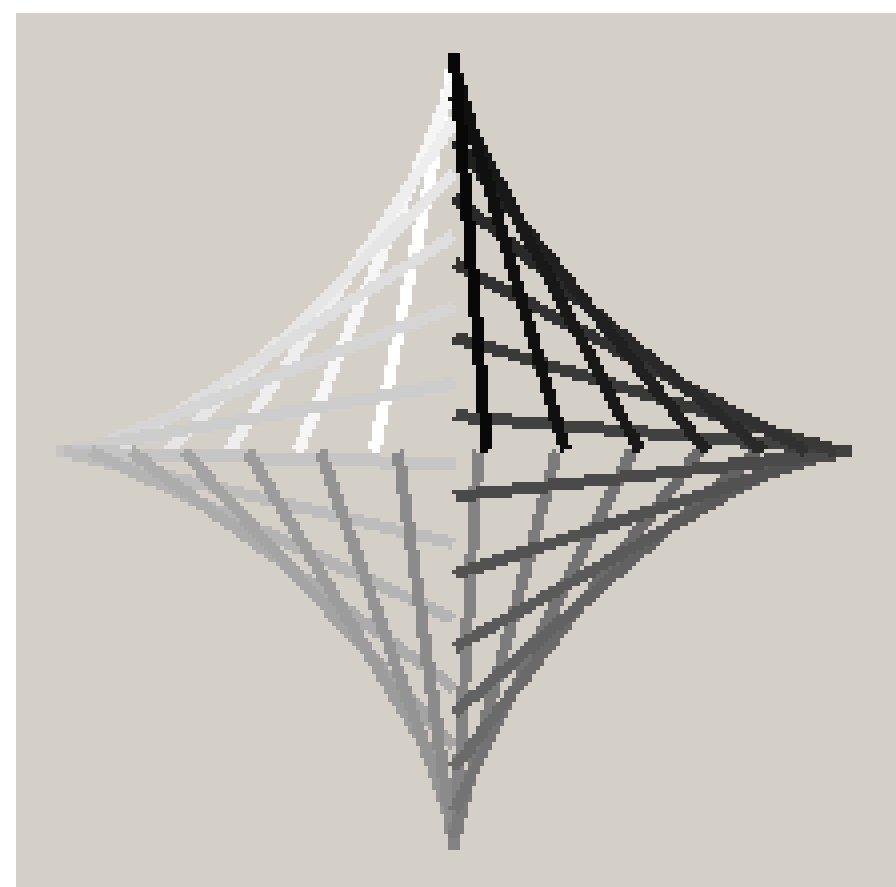

For a lever made of three links, simulation is sensitive to a pseudoinverse tolerance value:

```
* Unbalanced Lever
B1 (1,0) 1
B2 (-1,0) 2
C1 B1 (0,0)
C2 B2 (0,0)
C3 B1 B2
.tran 0 6.6 0.001
.options mpitol=1.0E-9
.plot CONSTRAINTS(C3)
.options gravity=1 FRAME=0.2s
*+ PLOTMODE=ANIMATEWITHSTROBO
```

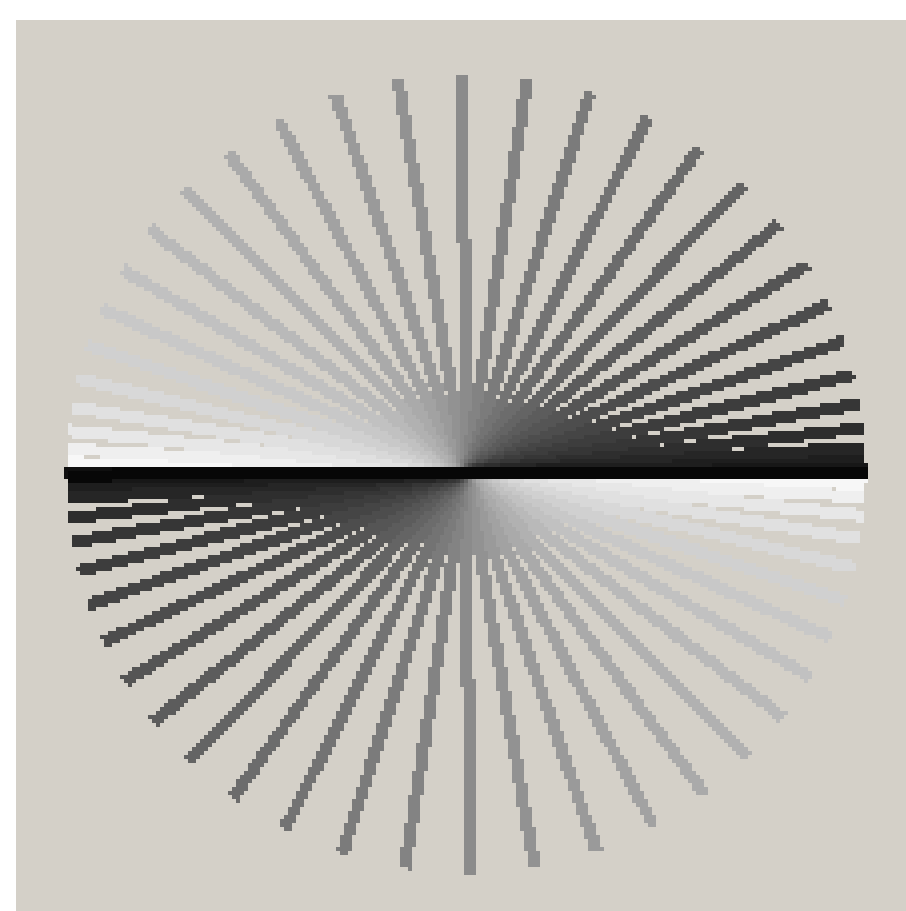